\documentclass[]{mn2e}
\usepackage{graphics}

\title[UIB simulation]{Simulation of UIB spectra with IR emission from CHONS molecules}
\author[R. Papoular]{Renaud Papoular$^{1}$\thanks{E-mail:papoular@wanadoo.fr}\\
$^{1}$Service d'Astrophysique and Service de Chimie Moleculaire,
CEA Saclay, 91191 Gif-s-Yvette, France}
\begin{document}

\date{Accepted . Received ; in original form }

\pagerange{\pageref{firstpage}--\pageref{lastpage}} \pubyear{2002}

   \maketitle
\label{firstpage}

\begin{abstract}

The present work purports to identify candidate carriers of the UIBs. This requires a procedure for the computation of the emission spectrum of any given candidate. The procedure used here consists in exciting the carrier into a state of internal vibration, waiting until the system has reached dynamic equilibrium and, then, monitoring the time variations of the overall electric dipole moment associated with this vibration. The emission spectrum is shown to be simply related to the FT of these variations. This procedure was applied to more than 100 different chemical structures, inspired by the exhaustive experimental and theoretical analyses of Kerogens, the terrestrial sedimentary matter, which is known to be mainly composed of  C, H, O, N and S atoms. From this data base, 21 structures were extracted, which fall in 4 classes, each of which contributes preferentially to one of the main UIBs. Summing their adequately weighted spectra delivers an emission spectrum which indeed exhibits the main UIB features (allowing for computational errors inherent in the chemical simulation code). By changing the weights, it is possible to change the relative band intensities so as to mimic the corresponding (moderate) changes observed in the sky. The defects of the present simulation are discussed, and directions for improvement are explored.

\end{abstract}


\section{Introduction}

The quest for a realistic model of the structure, composition and excitation of the carriers of UIBs (Unidentified Infrared Bands) has been going on for more than three decades now. An essential requirement is, for such a model, to correctly reproduce the spectral features of UIB 
emission, which, by now, are very well documented in a large variety of astronomical sites. Among the latest publications, note the exhaustive analysis of a rich collection of spectra from star-forming regions (Smith et al. 2007). Despite some variations from object to object, there are enough similarities that it has been possible to partition most of the observed spectra into a small number of classes (see, for instance, Peeters et al. 2002, 2004a; van Diedenhoven et al. 2004), thus allowing systematic comparisons to be made with laboratory or theoretical spectra. 
 
Most present day candidate model carriers are big molecules of widely different sizes (see Kwok and Sandford 2008, Tielens 2008, Kwok 2009), the size being limited, of course, by experimental or computational constraints,or even by theoretical considerations. While the experimental production and spectral analysis of single macromolecules have proven to be daunting, modern computing capabilities make it easier to use theory or chemical simulation for modeling purposes.

The most usual computational way of determining the IR (InfraRed) spectrum of a given structure (molecule or grain) is to perform a Normal Mode Analysis (NMA; see Wilson et al. 1955). This delivers all the different modes of atomic vibration of infinitesimal amplitude, characteristic of the structure. For a grain made up of N chemically bound atoms, there are 3N-6 such modes, each characterized by its frequency and the directions and velocities of the excursion of each atom from its equilibrium position along the 3 space coordinates. Further treatment yields the mode \emph {ir intensity} (or \emph{line intensity}, or \emph{integrated band intensity}), which is proportional to the \emph {absorbance} at the corresponding frequency, the quantity commonly measured in the laboratory. NMA can be performed using chemical simulation codes based on different theories and approximations: Molecular Mechanics (MM+, Ambers, BIO+, etc.) and quantum chemical codes: Semi-empirical (AM1, PM3, etc.), \emph{ab initio} (e.g. DFT) methods, which differ widely in accuracy and time consumption. These are available with commercial packages such as HyperChem, Spartan, Gaussian, etc.

As early as 1992, Brenner and Barker \cite{bre} warned that ``unambiguous identification of the IS (InterStellar) emitter cannot be established just on the basis of matching the emission frequencies to laboratory absorption spectra". For emission spectra differ in many respects from absorption spectra: the features of the former are generally shifted and broadened relative to those of the latter, because of anhormonicity effects, which set in as soon as the grain is excited; correlatively, they depend on the excitation process (photon absorption, chemical reaction, etc.). Brenner and Barker themselves computed the emission of excited benzene and naphtalene based on a) the thermal equilibrium assumption (statistical distribution of energy among molecular states), b) Einstein's spontaneous emission coefficients, as deduced from measured (relative) absorption coefficients at a few characteristic spectral frequencies of the two molecules (see also Allamandola et al. 1989).  Intrinsic bandwidth and underlying emission continuum were assumed. Several off-springs of this procedure have been used (e.g. Verstraete et al. 2001, Boersma et al. 2010, 2011); the state of the art in this domain is clearly summarized by Bauschlicher et al. \cite{bau}.

Addressing the same issue, Papoular \cite{pap01}, \cite{pap12b} proposed a distinct procedure which bypasses NMA altogether. Based on the unversal relation between emission, electric dipole moment and frequency, he showed that the emission spectrum of an excited molecule is simply related to the energy content and overall dipole moment of the structure. The proposed procedure hence consists in monitoring the variations in time of the energy content and overall dipole moment of a selected model structure after it has been excited. The FT (Fourier Transform) of these variations deliver the emission spectrum. As the specific anharmonicity of the molecule, and all its effects, are reflected in its internal motions, it also affects both the energy and dipolar spectra, and, hence, the emission spectrum. As a consequence, the \emph{absorption line} spectrum of NMA is replaced by an \emph{emission band} spectrum. No assumptions are added to shift or broaden the bands artificially. Plateaus arise naturally from overlapping band wings. If the excitation energy is high enough, overtone and combination bands automatically emerge from the analysis. 

The anharmonicity is generally built in the dynamics code used for chemical simulation. In semi-empirical codes, it is generated by a system of parameters tailored by analogy with measured properties of structures made of the particular atomic species and bonds of interest. The PM3 code used here is best fit to simulate hydrocarbon molecules. Semi-empirical and Molecular Mechanics commercial codes were used to demonstrate relaxation after excitation (see Papoular 2001 and 2002), as well as anharmonicity effects: line shift, mode-locking, combination frequencies, Fermi-resonance (see Papoular 2006) and line broadening (Papoular 2012). 

Both DFT (Density Functional Theory) and Semi-empirical quantum chemistry methods deliver accurate Normal Mode intensities and frequencies. However, DFT methods are basically designed to deal with \emph{equilibrium states}, so they are not directly applicable to the analysis of dipole moment variations. To my knowledge, even Time-Dependent DFT (TDDFT) has been applied only to adiabatic transitions between equilibrium states, not to \emph{dynamics}.

Although, the procedure described above provides a sound theoretical transition from NMA to emission spectra and yields a degree of line broadening, this does not allow to mimic the wide mid-IR bands observed in the sky, unless very large aggregates of atoms are considered, or the emission spectra of a very large number of different medium-sized molecules are summed up. The former option is excessively computer-time consuming, and is not pursued further here. As for the latter, it is preferably started with a survey of the Normal Mode spectra (easier to compute!) of structures built up with the most common chemically active atoms: C, H, O, N and S (CHONS for short). Even with this restriction, the number of possible combinations is daunting.  Papoular \cite{pap10} therefore sought cues in kerogens and coals, whose spectra have features in common with the UIBs, and which have been abundantly modeled using laboratory analysis and chemical codes (e.g. Durand 1980, Behar and Vandenbroucke 1986, Carlson 1992,  Speight 1994). It turned out that some types of structures have spectral lines falling mostly within one or two UIBs. Changing slightly the structure or composition slightly displaces the spectral lines, thus helping to build a model band. More than a hundred normal mode spectra were computed and distributed over eight families of structures, each contributing mainly to one or two spectral regions. By combining these spectra in the right proportions, it was possible to synthesize spectra bearing some resemblance with UIB spectra. 

These computational experiments led to the selection of a number of structural types exhibiting spectral features which appeared to be promising for our present purposes. In the present work, the \emph{emission} spectra of many more structures of these types were computed according to the procedure outlined above. Each was given a weight and their weighted sum compared with UIB spectra. The details and outcome of this endeavour are given below.

Section 2 lays down the theoretical basis of the proposed method of spectral emission calculation; it is complemented by Appendix A. The details of its implementation in the computational procedure are given in Sec. 3; Sec. 4 illustrates the 4 types of chemical structures that were included, by displaying one of each type, all the others being collected in Appendix B; Sec. 5 assembles the corresponding spectra in 4 graphs, one for each type, and compares the resulting overall emission spectrum with an observed galactic spectrum. Section 6 discusses the defects of this simulation and Sec. 7 explores possible improvements. 

\section{Radiated energy}

Present chemistry simulation codes do not include spontaneous radiation. A simple analytical argument can, however, be made to deliver an operational approximation to reality. When energy is deposited in a molecule, several modes are excited; which ones depends on the excitation process. If the modes were strictly harmonic and, therefore, completely uncoupled, each and everyone would start radiating in the IR, at its own rate, different than the others. Assuming the excitation rate to be slow enough, each mode would ultimately radiate the energy which was initially deposited in it; the radiated power would be obtained by multiplying this with the frequency of excitation events. But, in a real molecule above its ground state, anharmonicity cannot be overlooked and the modes are coupled more or less. As a consequence, intramolecular vibration redistribution then quickly changes the initial distribution into a new one, which is mostly independent of the excitation process, but characteristic of the molecule itself when in dynamical equilibrium (see Papoular 2001, 2002). 

Internal relaxation strongly impacts the spectral emission, as we now show on a simplified ideal case. Assume first, for simplicity, that the molecule has only 2 vibration modes, and let $E_{i}$ be the energy, $\tau_{i}$ the radiative lifetime of the i\emph{th} mode, with $i=1,2$, and $e$ the energy exchanged between modes. Also let $c_{i}^{-1}$ represent the fraction of $E_exc$ (the total excitation energy) that is stored in mode $i$ when internal relaxation is completed, i.e. the above mentioned characteristic energy distribution in dynamic equilibrium. Then,

\begin{eqnarray}                                            
\dot E_{1}= -E_{1}/\tau_{1}+\dot e \\
\dot E_{2}= -E_{2}/\tau_{2}-\dot e \\
\dot e=a(c_{2}E_{2}-c_{1}E_{1}),
\end{eqnarray}

where the dot designates the derivative and $a$ is the mode coupling constant. Note that, when $E_{i}\propto c_{i}^{-1}$, energy exchange stops. Internal and radiative relaxation may be represented by a solution to this set, of the form 

\begin{eqnarray} 
E_{1}=\alpha_{1}exp(-t/\tau)+\beta_{1}exp(-t/\sigma) \\
E_{2}=\alpha_{2}exp(-t/\tau)+\beta_{2}exp(-t/\sigma) \\
e=\alpha exp(-t/\tau)+\beta exp(-t/\sigma).
\end{eqnarray}

The meaning of these equations is clear when, as is usually the case, the coupling constant, $a$, is strong enough that $\sigma\ll\tau$. In a first, short, intramolecular relaxation phase (the transient, with time constant $\sigma$), radiation is insignificant, but the energies in the two modes are forced to quickly readjust (from $E_{0i}$ to $\sim\alpha_{i}$) so as to nearly annul $\dot e$ (eq. 3). They then can relax at the same, much slower, rate, $\tau$, intermediate between the two radiation rates $\tau_{1}$ and $\tau_{2}$, while energy is being continuously transfered from the less radiative to the more radiative mode.

That $\sigma\ll\tau$, results from the fact that the radiative lifetime in the IR is very long ($>$1ms), while the intramolecular vibrational relaxation is shorter than 1 ns, in general, as illustrated by numerical modeling on several molecules (see Papoular 2001, 2002 and 2006).

The 8 unknown constants appearing in the solution eq. 4 to 6 are determined by first substituting in eq. 1 to 3, and setting to 0 the sum of the coefficients of the exponentials in $\alpha$ and $\sigma$, separately and for each equation. Then, initial conditions are applied: $E_{i}(t=0)=E_{0i}$. The results are cumbersome to an unwarranted degree unless $\sigma\ll\tau$. In that case, the procedure leads to 

\begin{equation} 
\beta_{1}=-\beta_{2},
\end{equation}

Moreover, the intra-molecular and radiative phases hardly overlap, so that, at the start of the latter, there is no more energy exchange between modes:
$\dot e=0$, so that

\begin{equation} 
c_{1}\alpha_{1}=c_{2}\alpha_{2}\,.
\end{equation}

Then, the required constants are approximately given by

\begin{eqnarray} 
\alpha_{1}=c_{2}\frac{E_{01}+E_{02}}{c_{1}+c_{2}} \\
\alpha_{2}=c_{1}\frac{E_{01}+E_{02}}{c_{1}+c_{2}} \\
\beta_{1}=-\beta{2}=\beta=\frac{c_{1}E_{01}-c_{2}E_{02}}{c_{1}+c_{2}} \\
\tau=\frac{\tau_{1}\tau_{2}(c_{1}+c_{2})}{c_{1}\tau_{1}+c_{2}\tau_{2}} \\
\sigma=\frac{1}{a(c_{1}+c_{2})} \\
\alpha=\alpha_{1}\alpha_{2}\frac{\tau_{1}-\tau_{2}}{\alpha_{2}\tau_{1}+\alpha_{1}\tau_{2}}\,.
\end{eqnarray}

All of the above is borne out by numerical resolution of system 1 to 3. Finally, the energies radiated by modes 1 and 2 (after extinction of the transient) are, respectively

\begin{eqnarray} 
\int E_{1}dt/\tau_{1}=\tau \alpha_{1}/\tau_{1}=\frac{c_{1}+c_{2}}{c_{1}\tau_{1}+c_{2}\tau_{2}}\tau_{2}\alpha_{1} \\
\int E_{2}dt/\tau_{2}=\tau \alpha_{2}/\tau_{2}=\frac{c_{1}+c_{2}}{c_{1}\tau_{1}+c_{2}\tau_{2}}\tau_{1}\alpha_{2}\,,
\end{eqnarray} 
whose sum equals $\alpha_{1}+\alpha_{2}$, as it should, since there is no outlet for the initial energies other than radiation. The ratio of total energies radiated at the respective mode frequencies is 

\begin{equation}
\frac{\alpha_{1}}{\alpha_{2}}\frac{\tau_{2}}{\tau_{1}}=\frac{c_{2}}{c_{1}}\frac{\tau_{2}}{\tau_{1}}\,.
\end{equation}

The forms of the solutions above are easily generalized to any number of modes, assuming a uniform coupling constant. This is inocuous since the coupling constant only impacts the transient time constant, $\sigma$, which remains very short anyway (a few tens of picoseconds).

 The radiative lifetime is the inverse of Einstein's spontaneous emission coefficient (see Wilson et al. 1955)
\begin{equation}
A(\nu)=\frac{64\pi^{4}\nu^{3}}{3hc^{3}}m(\nu)^{2}\,,
\end{equation}
in electrostatic CGS units, and written in simplified form, for one particular transition and overlooking degeneracies. Here, $m(\nu)$ is the dipole moment oscillating component at frequency $\nu$. Now, it comes as no surprise (see Atkins 1997) that the corresponding IR intensity is also linked to
 $m(\nu)$ by
\begin{equation}
I(\nu)=2.6\nu m(\nu)^{2}\,,
\end{equation}
where $m$ is in Debye and $\nu$ in cm$^{-1}$. Thus, Einstein's coefficient scales like $\nu^{2}I(\nu)$ and $\nu^{3}m(\nu)^{2}$. Finally, the energy ultimately radiated by mode $\nu$ scales like

\begin{equation}
E^{*}\propto\nu^{3}c^{-1}(\nu)E_{exc}m(\nu)^{2}\,,
\end{equation}

where the mode index $i$ was replaced by its frequency. Equation 20 can also be derived from the classical oscillator emission by multiplying with the oscillator strength (see Kuhn 1962, Wilson et al. 1965). Its range of application is therefore wider than those based on Einstein's coefficient or integrated line absorption, as these are restricted to transitions from or to the ground state, and they are delta functions unaffected by anharmonicity, so they cannot account for line broadening and shift.

Appendix A below illustrates this procedure as applied to ethylene C$_{2}$H$_{4}$, whose sparse IR spectrum makes for a better presentation of the effects of anharmonicity, and differences between IR emission, absorption and energy distribution. \rm

\section{The computational procedure}
The present work is based on the use of various algorithms of computational organic chemistry, as embodied in the Hyperchem software provided
by Hypercube, Inc., and described in detail in their publication HC50-00-03-00, and cursorily, for astrophysical purposes, in 
Papoular \cite{pap01}. Here I use the
 improved version Hyper 7.5. While almost all the computations used standard algorithms, some required writing special purpose subroutines, a 
capability also provided by the software. Semi-empirical methods (PM3 and AM1) were used for molecules up to about 100 atoms in size; their accuracy for spectral frequencies is of a few percent.
Beyond that, the less accurate, but faster Molecular Mechanics (MM+) method was preferred. Anharmonicity is built in and tailored by comparison with laboratory experiments on hydrocarbon molecules. 

Once a molecule has been selected, the first step is the building and optimization of its structure, i.e. minimizing its potential energy 
by automatic standard procedure, at 0 K (no kinetic energy in the molecule). Next, a type of excitation is adopted. An isolated molecule in space can acquire energy by, for instance, absorbing a photon, or by one of its dangling (C-) bonds capturing an H atom, thus increasing its potential energy by an amount equal to the binding energy (Papoular 2012). In the following, we assume that the average interval between two successive excitation events is much longer than the IR lifetime of the molecule. So, the radiative \emph{power} of a molecule is the product of the frequency of exciting events and the \emph{energy} radiated by the molecule upon radiative relaxation. Although different excitation processes were studied, this paper focuses on H capture by one of the dangling C bonds that are likely to be present at any time at the molecular periphery.
 In this process, a free H atom approaches the grain closely enough to feel the attraction of a carbon atom whose bonds are incompletely occupied. 
Under favorable initial conditions, it falls into the corresponding potential well. It was shown that this capture reorganizes the distribution of 
the constitutive atoms in space so as to lower the ground state of
 the newly completed structure by an amount equal to the CH bond energy ($E_{b}\sim$4.5 eV). This is equivalent to depositing this energy in 
this new structure, an energy which is immediately and evenly divided into kinetic energy and potential energy (relative to the ground state of
the completed structure). The kinetic energy itself, initially localized in the CH stretching vibration, is then redistributed among
 this and other (not necessarily all) molecular vibrations. This could not occur were it not for the couplings between modes. While normal modes 
are uncoupled by definition, the vibration modes become coupled as soon as some energy is deposited in the molecule, because of anharmonicity, which sets in above the ground state. Note that, in this process, by contrast with UV photon excitation, the molecule remains in the \emph{electronic} ground state even though atomic vibrations are excited. 

To simulate this process on the computer, we start with an 
optimized structure \emph{at rest, with all bonds occupied}. One of the CH bonds is then rotated out of the molecular plane 
by several degrees and extended from its original length by a few hundredths of \AA{\ } so as to increase the molecular potential energy by, say,  
$E_{exc}$, which is tantamount to depositing this energy in the molecule. The molecule is then set free to perform a molecular dynamics run
 in vacuum, during which all atomic motions are monitored on the computer screen and recorded, as well as the total kinetic and potential 
energies, as a function of time. This first run thus simulates the first, transient, phase described by eq. 4 to 6.

Initially, as expected, the dominant motion is the CH stretch accompanied by excursions out of the molecular plane. Soon, however, the other atoms
 are dragged along: inspection and comparison of the variations of, say, the bond lengths immediately reveals that energy
 constantly flows from bond to bond, thus increasing the amount of energy shared by other bonds. This process translates into a continuously 
decreasing standard deviation, $\sigma$, of each
 and every variable, like the total kinetic energy of the system, for instance. After some time (the relaxation time), a \emph{dynamic equilibrium} 
sets in,which is characterized by constant average relative standard deviations. The relative standard deviation of the energy, $\sigma_{E}/E$ is roughly $N_{M}^{-1/2}$, where $N_{M}$ is the number of effectively excited normal modes (see Reif 1984). The relaxation time is highly variable, depending on the grain structure and size, and on its initial perturbation. It may be as long as hundreds of picoseconds. This is still much shorter than the minimum radiative lifetime ($\sim1$ ms). The energy radiated during this time is negligible.

At the start of this relaxation phase, the initial energy distribution among modes depends mainly on the excitation mechanism, but it tends asymptotically toward a different final distribution (the c's). I call this the "characteristic" distribution for it depends mainly on the molecular structure and composition.

In dynamic equilibrium, energy is still continuously exchanged between modes, but is essentially conserved in the molecule as radiation is not included in the PM3 code. The time averaged energy distribution can be identified with the $c^{-1}$'s of Sec. 2. Once this state has been reached, a new run is launched, to simulate the emission phase, during which all variables of interest are monitored. First, we focus on the 
instantaneous total kinetic energy. The FT of its excursions from its mean delivers the excitation energy spectrum $c^{-1}(\nu)E_{exc}$ required in eq. 20. It exhibits peaks in the vicinity of the normal modes. Over and above the physical broadening due to anharmonicity, these peaks are also artificially broadened by the finite length, $\Delta\,t$, of the run: this contribution to the overall width at half maximum (FWHM, cm$^{-1}$) is $\sim (c\Delta t)^{-1}$. The longer the run, the less significant will this contribution be compared with the physical broadening: in general, 10 ps will do.

The dipole moment of the molecule is similarly monitored in dynamic equilibrium, and its Fourier transform similarly gives the dipole moment spectrum
, $m(\nu)$, also required in eq. 20, which can finally be computed to give the desired emission spectrum.

\section{The selected model UIB carriers}

\begin{figure}
\resizebox{\hsize}{!}{\includegraphics{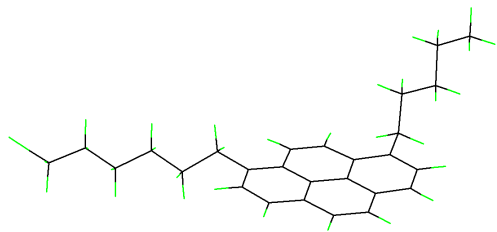}}
\caption[]{One member of the ``aromatics" class of selected structures. Atoms are not labeled, but C and H atoms are indicated by black and green bond segments, respectively. Aliphatic chains are expected to naturally bond to pure aromatics, and help slightly shift the spetral lines of the latter. Other members in Fig. 18.}
\end{figure}
             
\begin{figure}
\resizebox{\hsize}{!}{\includegraphics{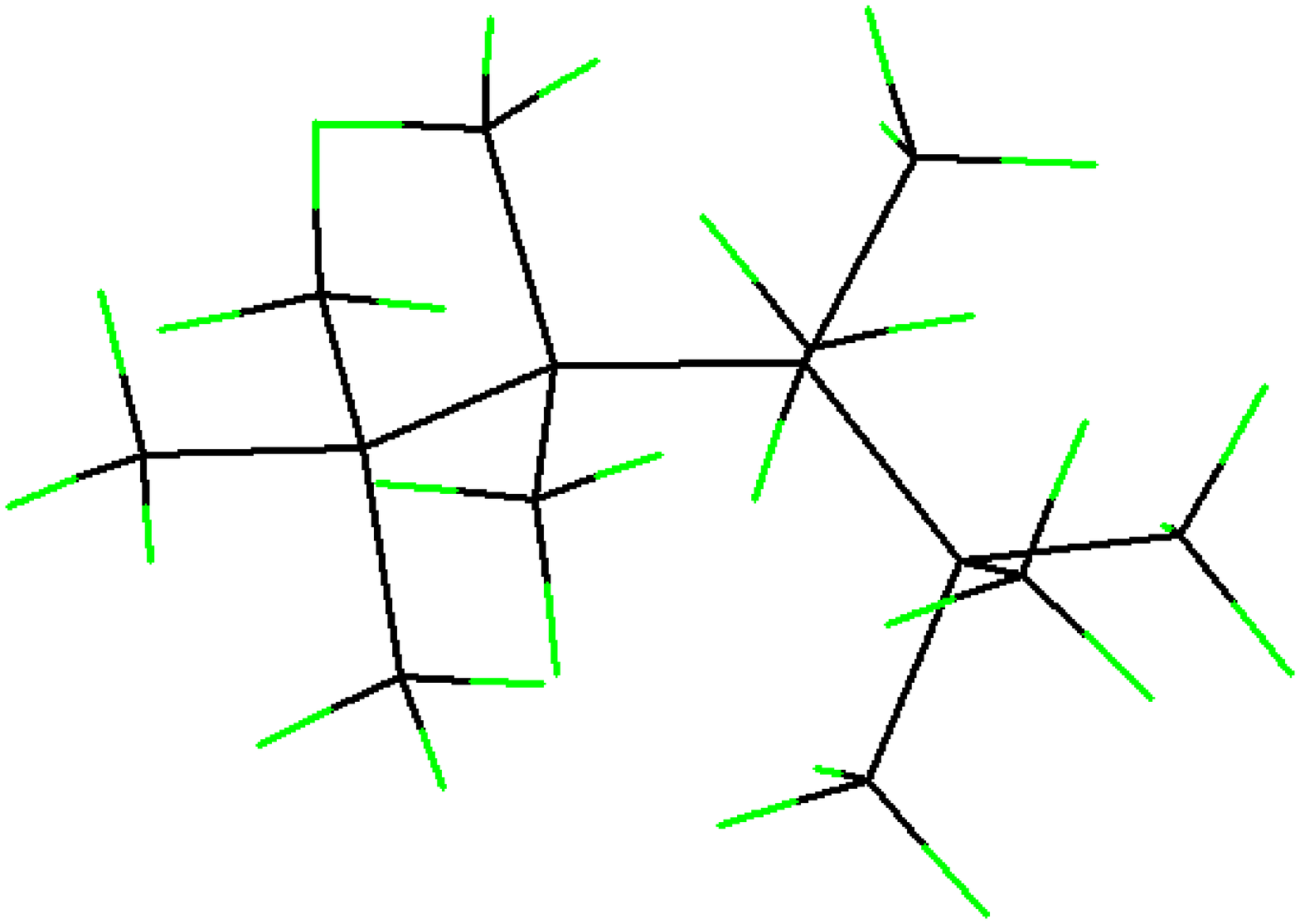}}
\caption[]{One member of the ``aliphatics" class. Another member in Fig. 19.}
\end{figure}

\begin{figure}
\resizebox{\hsize}{!}{\includegraphics{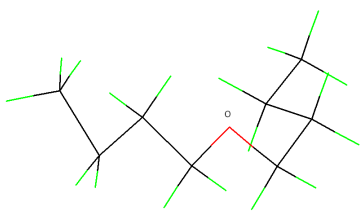}}
\caption[]{One member of the ``O-bridged chains". Oxygen is colored in red. Other members in Fig. 20 and 21.}
\end{figure}
\begin{figure}
\resizebox{\hsize}{!}{\includegraphics{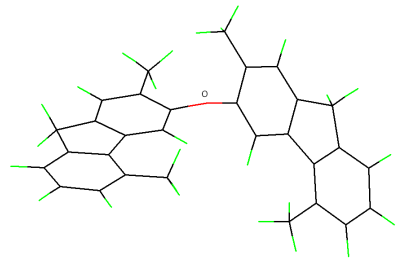}}
\caption[]{One member of the``trio" class. Other members in Fig. 22 to 24.}
\end{figure}

The attempt at simulating UIB spectra, as described in the present work, required the computation of emission from a large number of candidate structures so as to determine the characters of their possible contribution to UIB emission (see Papoular 2010). From this data base, 21 different molecules were selected, falling in 4 categories: aromatics, aliphatic chains, oxygen-bridged chains and ``trios" (essentially a 5-membered ring squeezed between two 6-membered rings). These denominations are used here in a very restrictive sense and apply only to the present, or slightly modified, structures. In this section, one member of each carrier category is illustrated (Fig. 1 to 4), The remaining structures are illustrated in a Appendix B (Fig. 18 to 24). 

The structure illustrated in Fig. 21 differs fundamentally from the others in that it is a cluster of four O-bridged chains (Fig. 3 and 20) bonded together by van der Waals forces. Its IR spectrum does not differ essentially from the sum of the spectra of its components. This suggests that the model UIB carriers proposed here could be taken separately or in clusters.

\section{The emission spectra}

\begin{figure}
\resizebox{\hsize}{!}{\includegraphics{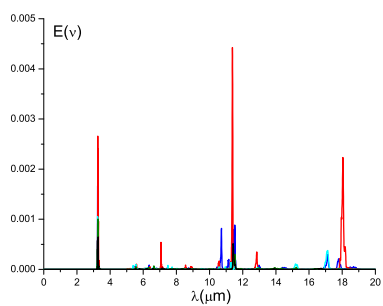}}
\caption[]{All 5 aromatic emission spectra.}
\end{figure}
\begin{figure}
\resizebox{\hsize}{!}{\includegraphics{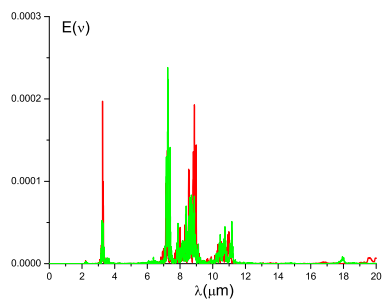}}
\caption[]{The 2 aliphatic emission spectra.}
\end{figure}
\begin{figure}
\resizebox{\hsize}{!}{\includegraphics{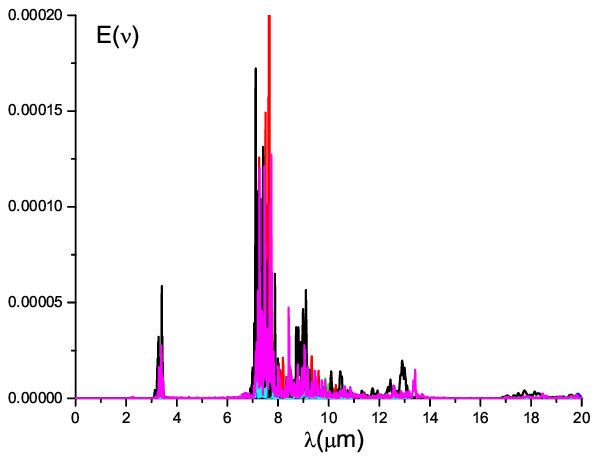}}
\caption[]{All 6 O-bridged-chain emission spectra.}
\end{figure}

Figure 5 to 8 each show the emission spectra of all members of one class, using one color for each member. 
The weighted spectra of the 21 selected molecules were added to give the expected overall emission. The weights were then tailored to steer the sum towards a typical observed and published galactic spectrum chosen for its quality and spectral extent (galaxy NGC1482; Smith et al. 2007). Figure 9 displays the result, together with the observed spectrum for comparison. Figure 10 shows the effect of smoothing the spectrum of Fig. 9. This simulates the expected effect of increasing the number of similar, but slightly different structures, of the selected set.

\begin{figure}
\resizebox{\hsize}{!}{\includegraphics{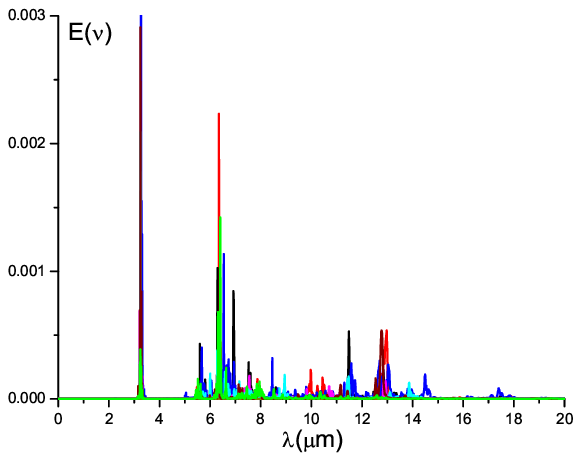}}
\caption[]{All 8 trio emission spectra.}
\end{figure}
\begin{figure}
\resizebox{\hsize}{!}{\includegraphics{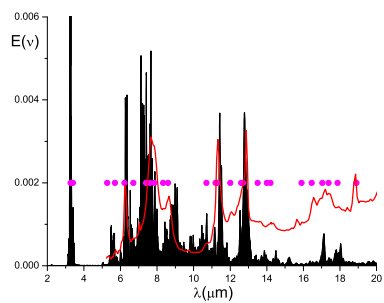}}
\caption[]{Black: the overall emission spectrum, obtained by adding all 21 member spectra, weighted by the coefficient $w$, tabulated in the last column of Table 1; area under curve is filled. Red curve: the flux curve, $I_{\nu}$, for NGC1482, from Smith et al. \cite{smi}; magenta dots mark the feature peaks tabulated in their Tab. 3; the continuum is not subtracted.}
\end{figure}

\begin{figure}
\resizebox{\hsize}{!}{\includegraphics{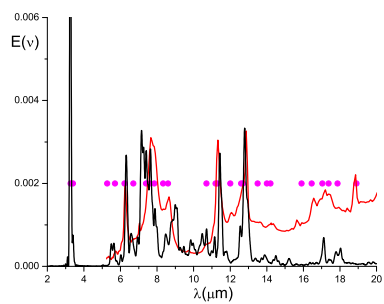}}
\caption[]{Black: FFT smoothing of the same overall emission spectrum as in Fig. 9; resolution: 0.1 $\mu$m . Red curve: the flux curve, $I_{\nu}$, for NGC1482, from Smith et al. \cite{smi}; magenta dots mark the feature peaks tabulated in their Tab. 3; the continuum is not subtracted.}
\end{figure}

\begin{figure}
\resizebox{\hsize}{!}{\includegraphics{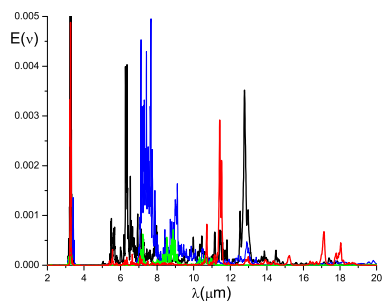}}
\caption[]{Breaking the overall emission spectrum by classes. Red: aromatics; green: aliphatics; blue: O-bridged chains; black: trios.}
\end{figure}

\begin{table*}[ht]
\caption[]{Atomic composition of dust}
\begin{flushleft}
\begin{tabular}{lllllllll}
\hline
Name & N$_{at}$ & C & H & O & S & $w$\\
\hline
Arom a & 59 & 27 & 32 & 0 & 0 & 3\\
\hline
Arom b & 36 & 24 & 12 & 0 & 0 & 0.2\\ 
\hline
Arom c & 41 & 27 & 14 & 0 & 0 & 1\\
\hline
Arom d & 31 & 19 & 12 & 0 & 0 & 1\\
\hline
Arom e& 44 & 22 & 22 & 0 & 0 & 3\\ 
\hline
Aliph Ch a & 44 & 14 & 30 & 0 & 0 & 2.5\\
\hline
Aliph Ch b & 53 & 17 & 36 & 0 & 0 & 2.5\\
\hline
O-bridge a & 27 & 8 & 18 & 1 & 0 & 12\\
\hline
O-bridge b & 27 & 8 & 18 & 1 & 0 & 20\\
\hline
O-bridge c & 72 & 8 & 18 & 1 & 0 & 20\\
\hline
O-bridge d & 30 & 9 & 20 & 1 & 0 & 20\\
\hline
O-bridge e & 33 & 10 & 22 & 1 & 0 & 10\\
\hline
O-bridge f & 111 & 33 & 74 & 4 & 0 & 20\\
\hline
Trio a & 57 & 30 & 26 & 1 & 0 & 1.5\\
\hline
Trio b & 52 & 30 & 20 & 1 & 1 & 0.5\\
\hline
Trio c & 112 & 59 & 50 & 3 & 0 & 2\\
\hline
Trio d & 54 & 31 & 22 & 0 & 1 & 1.5\\
\hline
Trio e & 52 & 30 & 20 & 1 & 1 & 1.5\\
\hline
Trio f & 60 & 31 & 20 & 1 & 1 & 2\\
\hline
Trio g & 58 & 31 & 26 & 1 & 0 & 40\\
\hline
Trio h & 60 & 31 & 28 & 1 & 0 & 5\\
\hline
Total & 7506 & 2377 & 4823 & 302.5 & 3.5 & 1\\
\hline
\end{tabular}
\end{flushleft}
\end{table*}

\section{Discussion}

It must be stressed again that the emission spectra displayed here are naturally broadened by the inherent anharmonicity of the molecular dynamics and the ensuing coupling between atoms, as described by the computational codes used in this work. The only artificial broadening involved in the calculation is due to the finite length in time of the molecular dynamics run, from which the dipole moment spectrum is deduced by FFT. For a 10 ps run, the broadening is 0.0035 $\mu$m, which does not impact the spectra notably.

When assessing the UIB simulation, it must be remembered that the inevitable defects of the chemical simulation codes entail an error of a few percent on the computed frequencies. With this in mind, it appears that the synthetic model spectrum exhibits a limited number of bands falling respectively at 3.25, 5.5, 5.65, 6.3, 7.4, 9, 11.4,12.8, 17.1 and 18.1 $\mu$m, near the main generic UIBs: (3.3,) 5.3, 5.7, 6.2, 7.7, 8.6, 11.3, 12.7 and 17-18 $\mu$m. The largest discrepancies occur for the 7.7 and 8.6 $\mu$m bands, and are of order 5 $\%$; they overemphasize the dip in between.

In Fig. 11, each colored line represents the sum of the weighted spectral contributions of the members of one class of structures. With the exception of the CH-stretching peak, each peak appears with a different color. This confirms that each of the 4 types of structures mainly contributes to a different UIB band, or set of bands:

-aromatics: 11.3, 17-18 $\mu$m\\
-aliphatics: 8.6 $\mu$m\\
-O-bridged chains: 7.7 $\mu$m\\
-trios: 5.3-5.7, 6.2, 12.7 $\mu$m.

These correlations stem from specific atomic motions favored by the structures. More detailed assignments are given in Papoular \cite{pap10} and were used to select the 21 structures considered here. These findings help tailoring new models to fit other observed spectra, which may differ in their relative band intensities. An interesting case in point is M82, in the halo of which, the ratios of the 3.3, 3.4 and 6-9 $_mu$m bands were found to vary significantly with distance from the Center (see Yamagishi et al. 2012).

Table 1 gives the total number of atoms, the number of C, H, O and S atoms in each molecule, and its weight, $w$, in the sum. The last row shows that, for this combination of molecules,

[O]/[C]=0.13, [H]/[C]=2, [S]/[H]=0.0015.

Not unexpectedly, these numbers correspond, in Van Krevelen's diagram, to kerogens type I (see Durand 1980, p122-126). Such kerogens are the least deeply buried in earth and, hence, the least aromatized (or mature, or ``evolved" or ``coalified") ones. Structural models for these, and other kerogens, were developed long ago; as stated in the Introduction, these models helped select the emitter structures considered in the present work.

Structural models developed for kerogens and coals include of course many more types of structures. For our present purposes, the choice is further constrained by astronomical data. Thus, alkene chains were not included, as double CC bonds carry strong features at 5.27 and 5.7 $\mu$m, whereas the corresponding features observed in space are relatively quite weak. Similarly, cyclohexanes (saturated 6-membered rings) could not be put to use to any extent.

Note that nitrogen and OH tails were used abundantly in modeling the 21- and 30-35-$\mu$m features of protoplanetary nebulae (Papoular, 2011). They were not used here, as these features do not show up in the spectrum of NGC1482 published by Smith et al. \cite{smi}. These authors also indicate that the conspicuous 18.9-$\mu$m band is mainly due to the ion SIII, which justifies its absence from the synthetic spectrum of Fig. 9.

One of the most conspicuous discrepancies between the synthetic and observed spectra is that the envelopes of the synthetic (black) ``bands" in Fig. 9 are not as smooth as those of NGC1482 (red). That is made clearer by zooming on the main bands, as in Fig. 12 to 15. Obviously, this is due to the sparseness of the selected model structures: the total number of atoms in the 21 selected molecules is 1068, giving rise to only 3198 vibration modes. It is clear that very many more structures can be built along the lines suggested by Fig. 18 to 24, to try and match the infinite diversity to be expected in space. But this must be left for later work. A glimpse of the effect this would have on the spectrum of Fig. 9 is gained by smoothing, as in Fig. 10. 

This is also a reminder of the fact that other selections of model structures may well fit observed spectra: in the present model, they are probably all present in space.

The close up views of the model bands reveal that the bunching of many peaks within a given band is necessarily accompanied by the appearance of underlying plateaus, formed by superposition of their wings. Also note that, between 6 and 15 $\mu$m, the sum spectrum never falls to zero, meaning that the selected set of molecules contribute to the underlying continuum. According to this model, therefore, the underlying continuum due to other families of carriers must not necessarily be drawn through the troughs between bands.

\begin{figure}
\resizebox{\hsize}{!}{\includegraphics{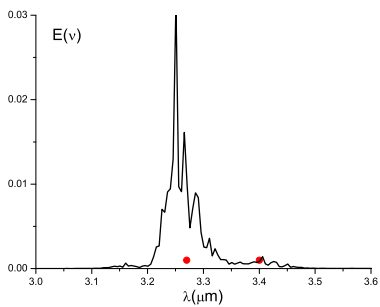}}
\caption[]{The 3.3 $\mu$m region.}
\end{figure}

\begin{figure}
\resizebox{\hsize}{!}{\includegraphics{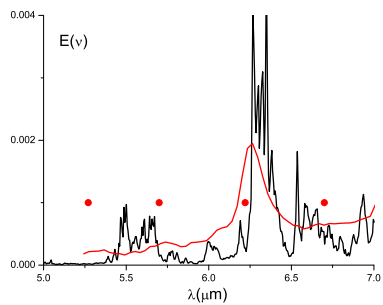}}
\caption[]{The 5.5-and 6.2-$\mu$m region.}
\end{figure}

\begin{figure}
\resizebox{\hsize}{!}{\includegraphics{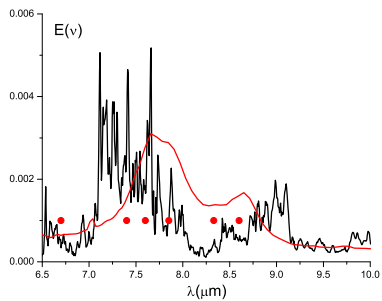}}
\caption[]{The 7.7-and 8.6-$\mu$m region.}
\end{figure}

\begin{figure}
\resizebox{\hsize}{!}{\includegraphics{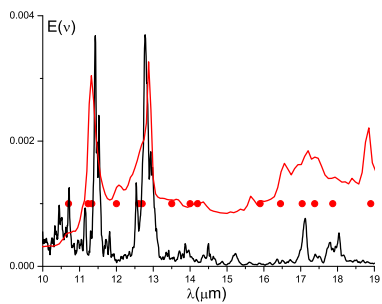}}
\caption[]{The 11.3-, 12.7- and 18- 
$\mu$m region.}
\end{figure}

\begin{figure}
\resizebox{\hsize}{!}{\includegraphics{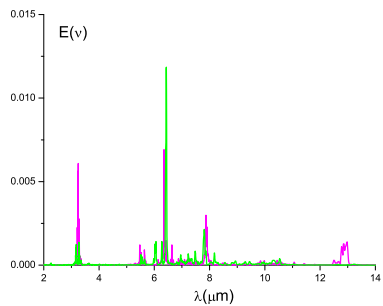}}
\caption[]{The effect of replacing H terminations of trio a (mauve) by methyl terminations (green): the ``3.3"-$\mu$m band is replaced by much weaker massifs on its wings, and the ``6.2"-$\mu$m band is considerably increased.}
\end{figure}

Another defect of the model is the excessive relative intensity of the 3.3-$\mu$m band: it is about 5 times more intense than the 7.7-$\mu$m band. Although it is not included in the NGC1482 spectrum for comparison, other observations suggest that the intensity of this band is of the same order as the 6.2- and 7.7-$\mu$m bands (see Peeters et al. 2004b). In the model spectrum, the main contribution to this band is that of the trio class, suggesting that this could be remedied by replacing some of the H terminations by methyls (or CH$_{2}$ chains).
A new structure was therefore built from trio a (Fig. 4), by replacing all the H's by CH$_{3}$'s (except those which cap the two pentagons). Their emission spectra are compared in Fig. 15: as expected, the CH stretchings are replaced by symmetric (umbrella) stretching of the methyls (to the left, probably because of code inaccuracy) and , to the right, symmetric stretchings of the two CH's capping a pentagon, and asymmetric stretchings of methyls. For coexisting H and CH$_{3}$ terminations, these aliphatic features could conveniently account for the 3.4-$\mu$ massif as well as the underlying plateau observed in the sky. Further modeling requires more simultaneous observational data on the near- and mid-IR domains.

With one exception (Fig. 19), the carriers considered here are all chemically bonded, medium sized structures. Molecules in space must be expected to suffer mild encounters and coalesce under Van der Waals attractive forces, to give clusters like in Fig. 19, which may become progenitors of dust grains. The study of such clusters should help assess the relative contributions to UIBs, of clusters (grains) and free molecules.

\section{Conclusion}
Clearly, more work is needed to obtain a satisfactory fit of model to observed spectra. The straightforward part might consist in including many more structures similar to those of the 4 mentioned classes. Exploring larger structures would help, as diversity increases with size, but this is more challenging. Also, that would call for more accurate and efficient codes and machines.

Studying the transition from molecules to grains, however, cannot be avoided if important issues are to be addressed, such as the emission continuum, the graphitization at the origin of the 2175 \AA{\ } extinction band and the evolution from the aromatic 3.3-$\mu$m to the aliphatic 3.4-$\mu$m band.

Even in the realm of medium-sized molecules, it should be possible to consider more tightly knitted structures than those studied here, which are essentially 2D. By contrast, taking on ionized molecules at present seems very laborious and uncertain with available simulations codes for molecular dynamics.

From the strictly chemical point of view, it would be interesting to understand why and under what circumstances, structures similar to those described in this work would come to be privileged.

Laboratory experiment could help advance these issues. In particular, the IR emission of a gaseous assembly of medium-sized molecules could be collected and spectrally analyzed at different pressures and temperatures: this may provide both a test of emission theories and an understanding of molecule-to-grain transition.

\section{Appendix A: line broadening}

\begin{figure}
\resizebox{\hsize}{!}{\includegraphics{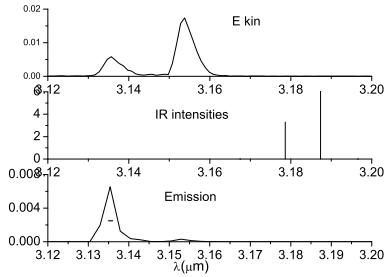}}
\caption[]{The CH stretching band of ethylene, obtained by the procedure outlined in Sec. 2. From bottom to top: radiated energy (or emission); normal modes (only 2 are IR active); energy distribution among modes.}
\end{figure}

 Figure 17 is an example of spectral results obtained by the procedure outlined in Sec. 2. It is a zoom on the CH stretching band of the ethylene molecule after excitation by bond extension (which injects 2.5 Kcal/mol in the molecule). The bottom graph represents the spectrum of the energy or power radiated by the molecule. This was obtained by monitoring the total electric dipole moment of the molecule and applying eq. 20.  For comparison, the middle graph represents the normal modes of ethylene, and the upper curve is the FT of the excursions of the kinetic energy from its mean, and represents the spectral distribution of energy in the excited molecule. Most normal mode lines are conspicuously absent in emission , because the corresponding vibration modes are either IR-inactive or not excited. As a consequence of excitation and anharmonicity, the separation between the two IR-active CH-stretching lines is nearly twice that of the corresponding normal modes, and the shorter wavelength emission peak is dominant. The horizontal segment, of length 0.0012 $\mu$m (bottom graph), represents the artificial broadening due to the finite calculation run time (30 ps). The real broadening, due to anharmonicity and mode coupling, is $\sim$0.004 $\mu$m. The relative shift of $\sim0.035 \mu$m (1 $\%$) between the middle spectrum and the other two is probably to be ascribed to minor inconsistencies between the two distinct algorithms which compute normal modes and molecular dynamics, respectively.

\section{Appendix B: Selected molecules}

\begin{figure}
\resizebox{\hsize}{!}{\includegraphics{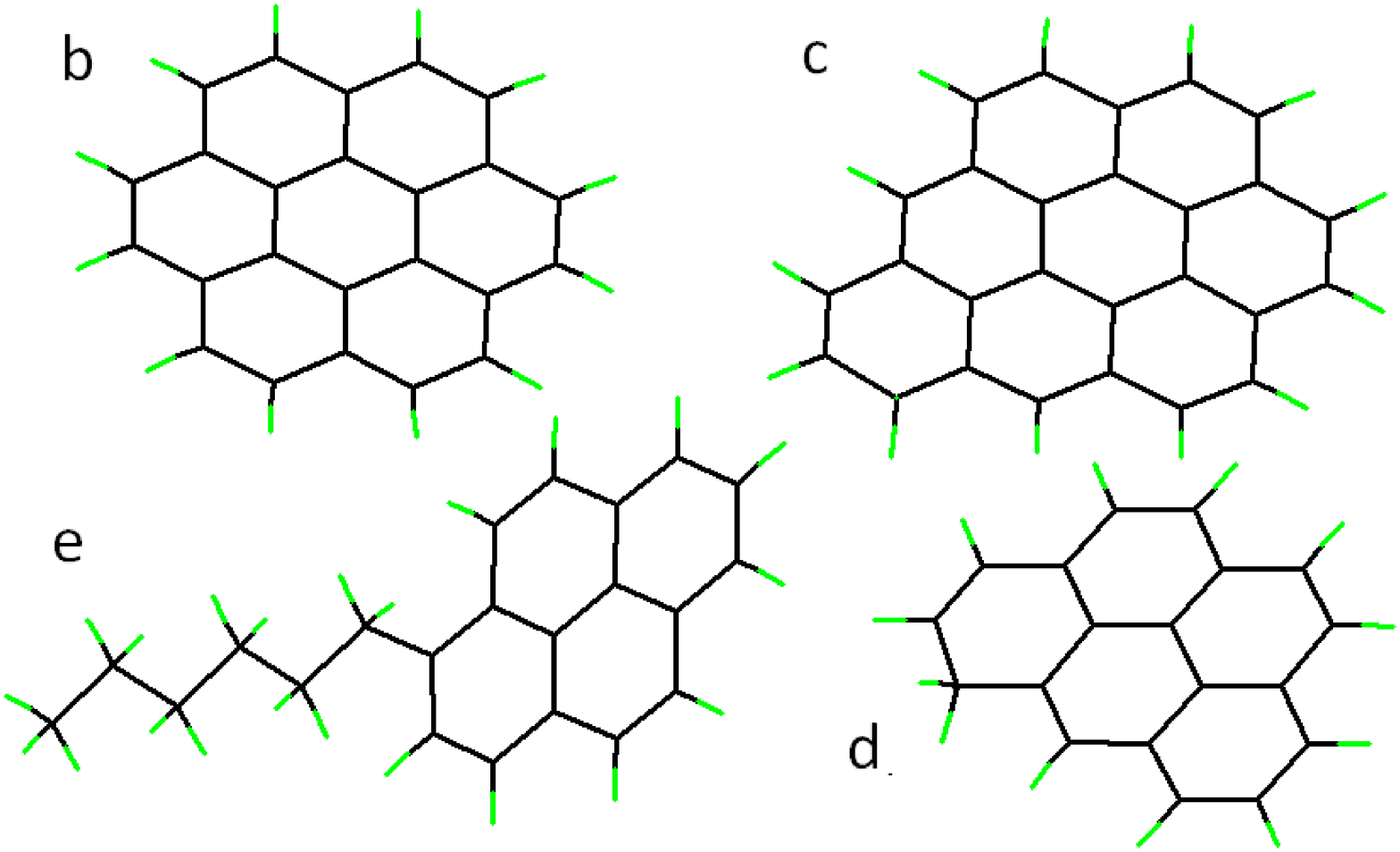}}
\caption[]{Other members of the aromatics family.}
\end{figure}

\begin{figure}
\resizebox{\hsize}{!}{\includegraphics{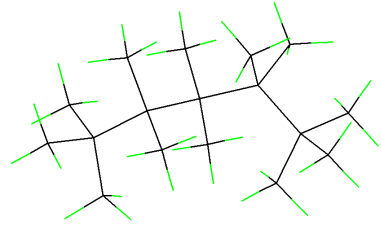}}
\caption[]{The other member of the aliphatics family.}
\end{figure}

\begin{figure}
\resizebox{\hsize}{!}{\includegraphics{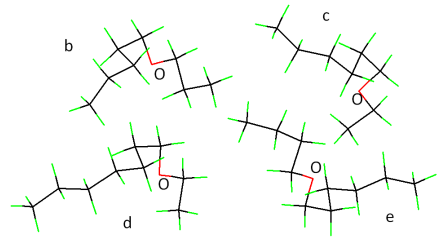}}
\caption[]{Other members of the O-bridged chains family.}
\end{figure}

\begin{figure}
\resizebox{\hsize}{!}{\includegraphics{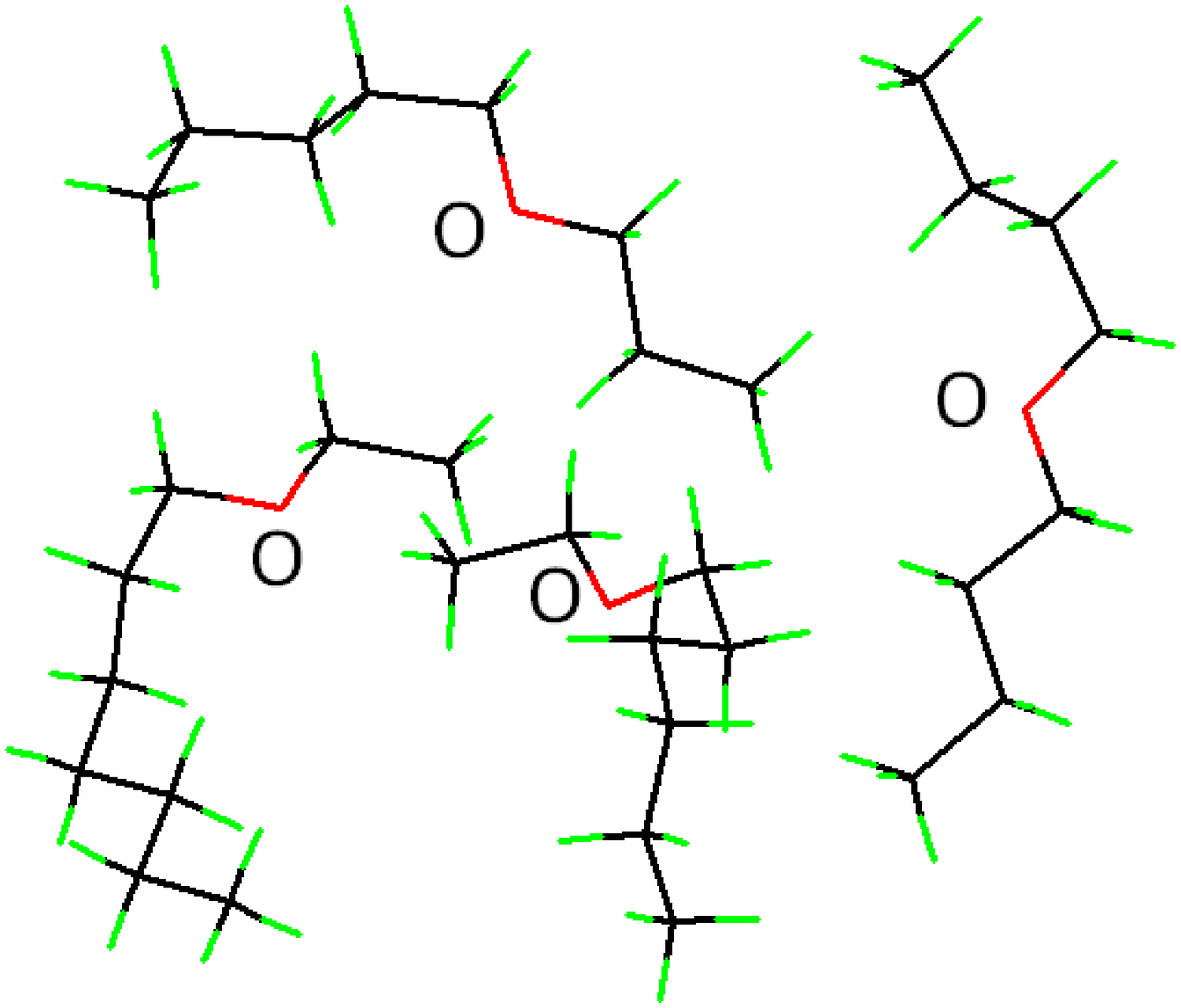}}
\caption[]{This last member of the O-bridged chains family is a cluster of chemically bonded molecules, held together by van der Waals forces.}
\end{figure}

\begin{figure}
\resizebox{\hsize}{!}{\includegraphics{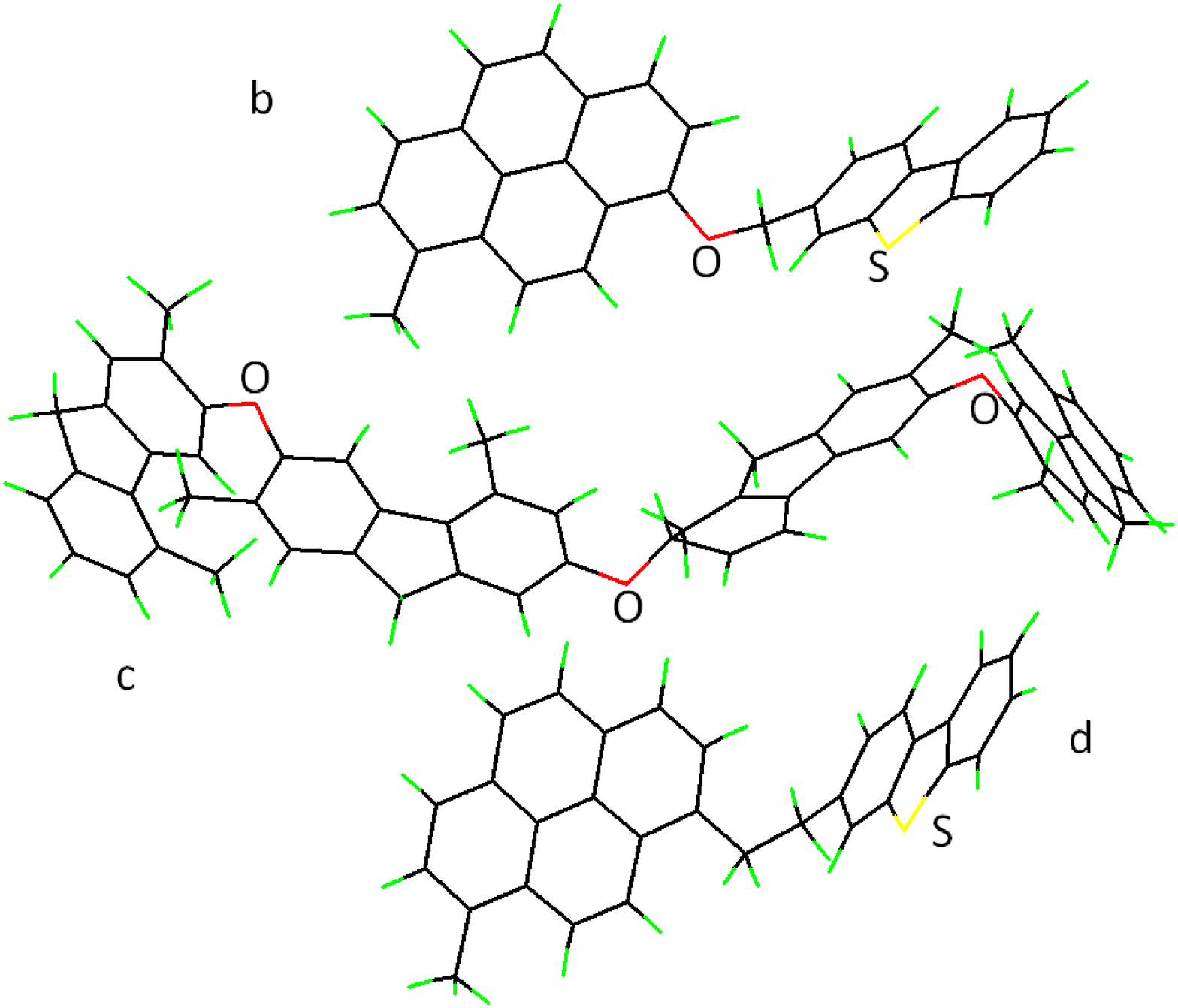}}
\caption[]{Other members of the trios family.}
\end{figure}

\begin{figure}
\resizebox{\hsize}{!}{\includegraphics{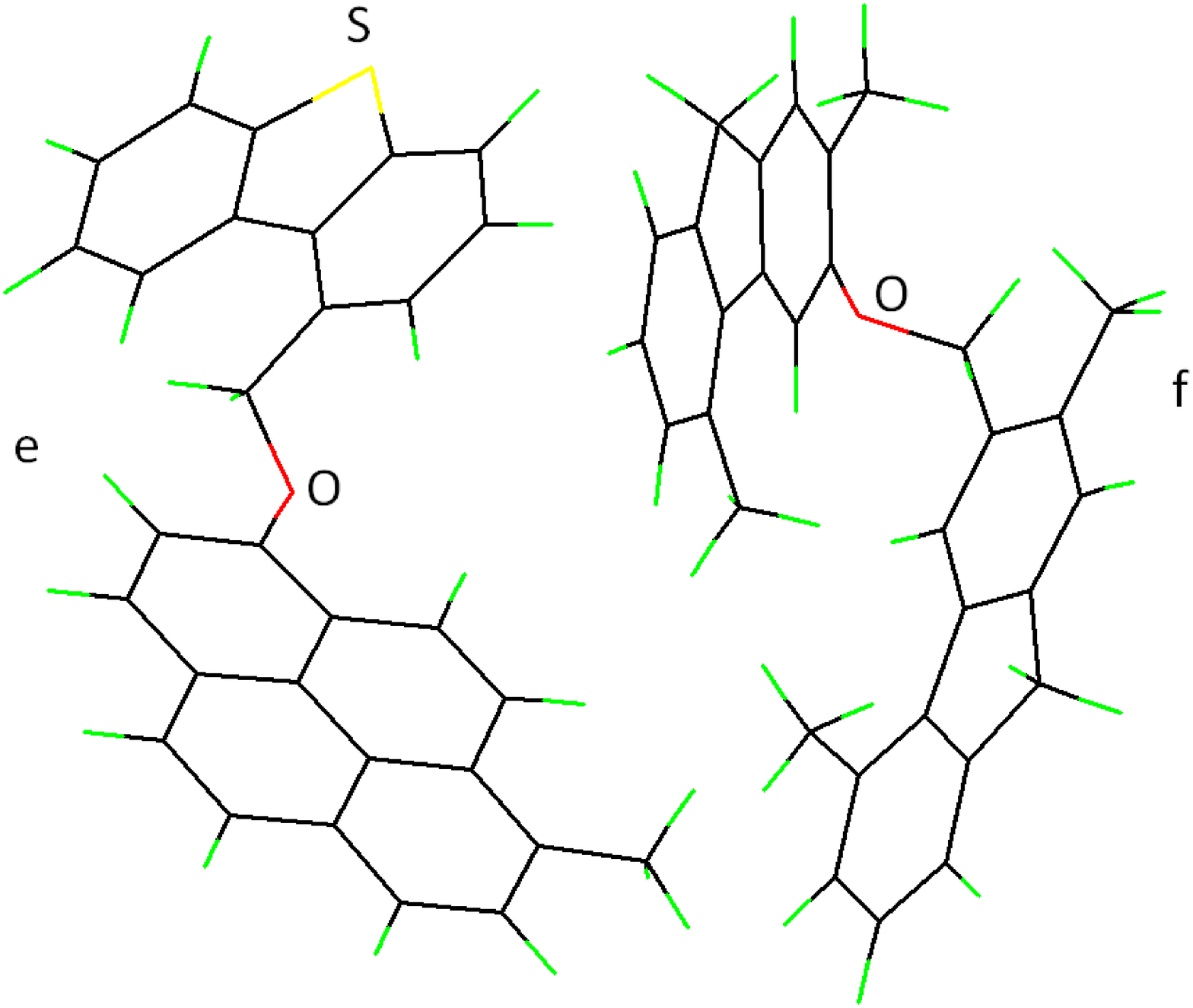}}
\caption[]{Other members of the trios family.}
\end{figure}

\begin{figure}
\resizebox{\hsize}{!}{\includegraphics{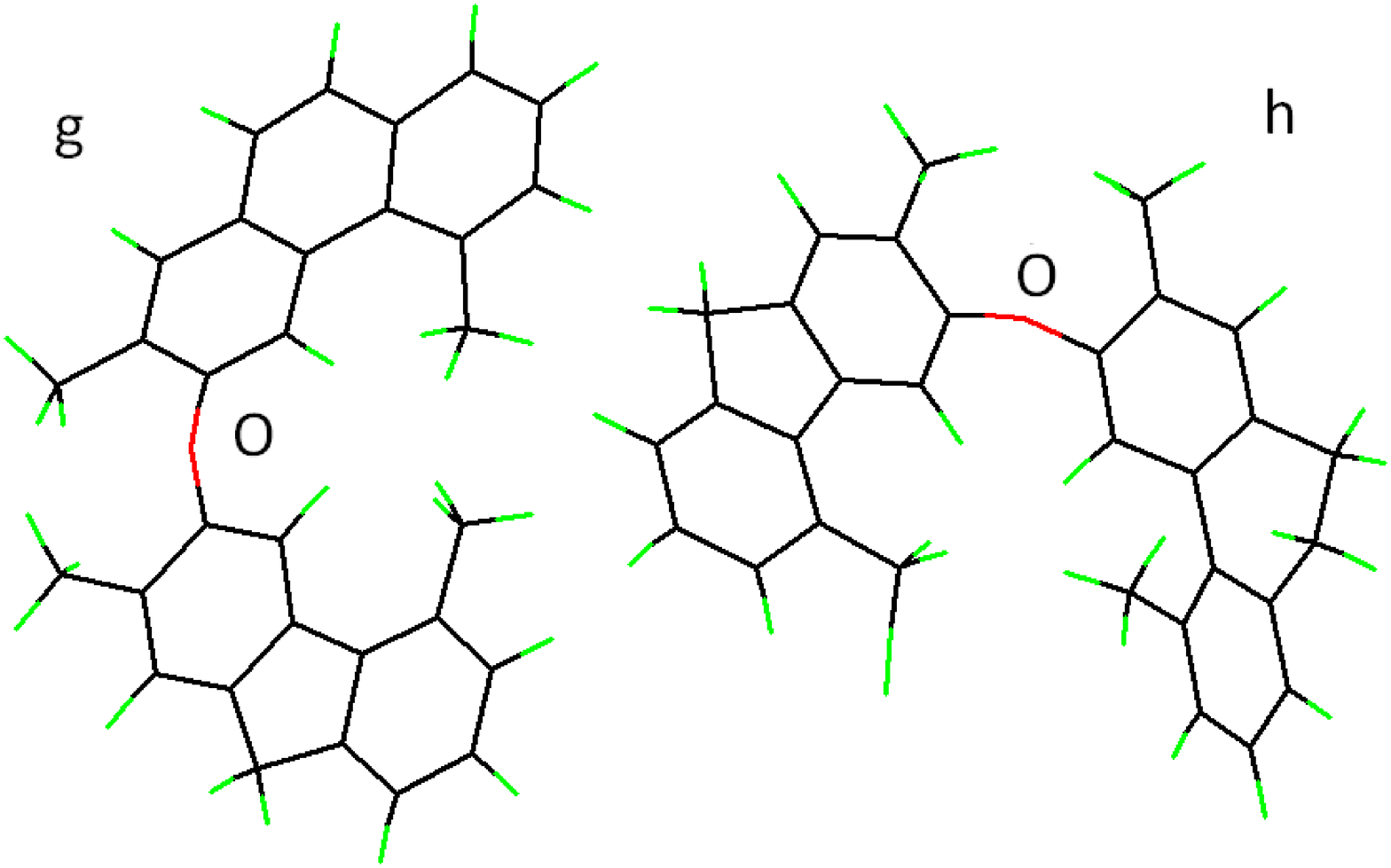}}
\caption[]{Other members of the trios family.}
\end{figure}

Figures 18 to 24 display all the chemical structures used used in the present work to model the spectrum of NGC 1482, except the head members of each class, displayed in Fig. 1 to 4.

\end{document}